\documentclass[referee]{raa}            

\usepackage{graphicx,times}             
\usepackage{natbib}
\usepackage{amssymb,amsmath}
\usepackage{adjustbox}
\usepackage{float}
\usepackage{makecell}
\bibpunct{(}{)}{;}{a}{}{,}

\usepackage[pagebackref=true]{hyperref}

\begin{document}

  \title{\textbf{Mock Observations for the CSST Mission: Main Surveys--the Stray Light} }

   \volnopage{Vol.0 (20xx) No.0, 000--000}      
   \setcounter{page}{1}          

   \author{Xian Jing-Tian 
   \inst{1,2}
   \and Lin Lin \inst{3}
   \and Fang Yue-Dong \inst{4}
   \and Zhang Xin \inst{5}
   \and Xu You-Hua \inst{5}
   \and Meng Xian-Min \inst{5}
   \and Tian Hao \inst{5}
   \and Zhang Tian-Yi \inst{1}
   \and Ban Zhang \inst{1}
   \and Li Guo-Liang \inst{6}
   \and Xu Shu-Yan \inst{1,2}
   \and Wang Wei \inst{1,2}
   }

   \institute{Changchun Institute of Optics, Fine Mechanics and Physics, Chinese Academy of Sciences, Changchun 130033, China; {\it wangwei123@ciomp.ac.cn}
   \and University of Chinese Academy of Sciences, Beijing, China
   \and Shanghai Astronomical Observatory, 80 Nandan Road, Shanghai, China
   \and Universit\"ats-Sternwarte M\"unchen, Fakult\"at f\"ur Physik, Ludwig-Maximilians-Universit\"at M\"unchen, 81679 M\"unchen, Germany
   \and National Astronomical Observatories, Chinese Academy of Sciences, Beijing 100101, China
   \and Purple Mountain Observatory, Chinese Academy of Sciences, Nanjing 210023, China
   }

\abstract{Stray light significantly influences the detection capabilities of astronomical telescopes. The actual stray-light level during observations depends not only on the telescope’s inherent stray-light suppression capability but also on its operational orbit conditions. Accurate estimation of stray-light levels is crucial for assessing image quality and performing realistic scientific simulations. To rapidly estimate stray-light levels under realistic, complex operational conditions, we developed an analytical model tailored to the China Space Station Telescope (CSST). Our model simulates stray-light backgrounds generated by off-field sources such as moonlight, starlight, and earthshine, incorporating the effects of zodiacal light, as well as scattering and ghost images induced by bright in-field stars. The proposed method allows quick and accurate evaluation of stray-light conditions, facilitating both image simulation and observational scheduling. }
$\cdots\cdots$
\keywords{Stray Light; Astronomical Space Telescope; Observational Scheduling; Image Simulation}

   \authorrunning{Xian Jing-Tian \& Wang Wei }            
   \titlerunning{CSST Main Survey Simulator: the Stray Light}  

   \maketitle

%
%
\section{Introduction}           
\label{sect:intro}

Stray light refers to unwanted radiation reaching a telescope’s detector \citep{Fest2013} from various sources, including sunlight, moonlight, zodiacal light, and earthshine \citep{Clermont2024,Boyd2022,Johnson2011}. Such undesired illumination can significantly degrade image quality by increasing background noise, reducing contrast, blurring celestial targets, and introducing spurious signals. Consequently, faint astronomical objects can become obscured, potentially causing incorrect detections or classifications \citep{Kahan2013,Smith2000,Kahan2019}. 

For space telescopes conducting survey missions, precise estimation of stray light, particularly from off-field sources, is essential for optimizing observational efficiency given their limited operational lifetimes. The China Space Station Telescope (CSST), with its wide 1.1 square-degree field of view \citep{Zhan2021}, inevitably includes bright stars during observations \citep{Chabot2023,Krist2023}. These stars can induce blurring and ghost images through scattering by optical surfaces and reflections between the filters and detectors \citep{Peterson2004}. Therefore, accurate modeling of both off-field and in-field stray light distributions is crucial for ensuring high-quality observational data. 

The stray light irradiance and its spatial distribution vary significantly with operating conditions \citep{Johnson2017,Bruce2015}. Critical factors include the telescope's pointing direction relative to the Sun, lunar phases, and angles to Earth's bright limb. CSST has established preliminary conservative constraints to mitigate stray-light contamination, maintaining minimum avoidance angles of 50° from the Sun, 40° from the Moon, 80° from Earth’s bright limb, and 30° from Earth’s dark limb. 

These constraints, however, require further refinement. For instance, the lunar avoidance angle currently relies on the brightness of the full Moon, yet the brightness difference between new and full Moon can reach 8–10 magnitudes. Earthshine conditions are similarly complex, influenced by both the solar position and Earth’s bright limb angle. When the angle between the line-of-sight and the Sun ranges approximately between 50° and 65°, adjustments to the aperture door orientation are necessary, potentially altering earthshine levels. Furthermore, celestial bodies within the solar system—such as Venus, Mars, and Jupiter—as well as bright stars, also serve as secondary stray-light sources. At viewing angles between approximately 50° and 80° from Earth's bright limb, extending exposure time can improve the signal-to-noise ratio, although the specific duration must be carefully chosen based on real-time stray-light levels. 

\begin{figure}[htbp]
    \centering
    \includegraphics[width=0.75\linewidth]{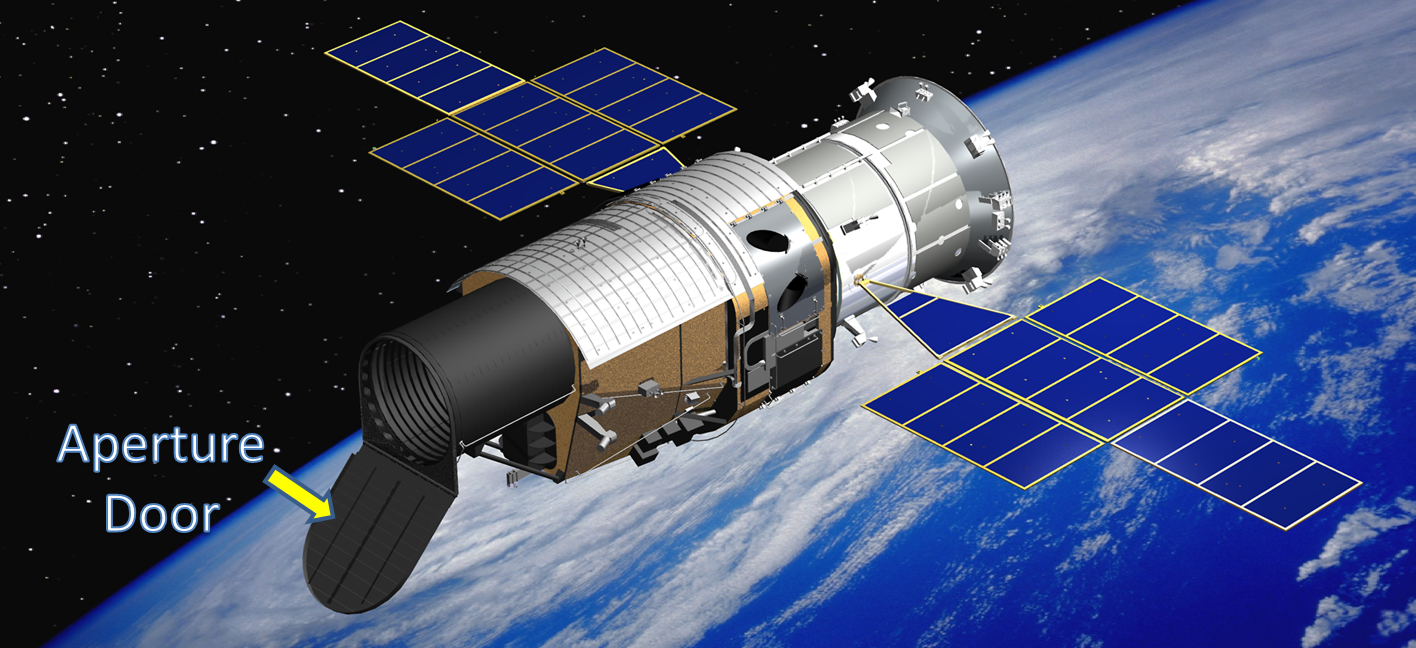}
    \caption{CSST and its aperture door \citep{Zhan2021}}
    \label{fig:CSST and its aperture door}
\end{figure}

Currently, the mainstream stray-light analysis methods for astronomical telescopes are ray-tracing method and experimental method. The ray-tracing method involves detailed modeling of internal optical structures and external sources using commercial optical design software (e.g., Zemax, Code V, FRED). It traces the propagation paths of rays within the system to analyze the mechanisms, intensities, and spatial distributions of stray light. It's capable of accurately modeling complex optical structures providing intuitive visualization of stray-light propagation paths. The experimental method involves building actual optical systems or scaled models in a laboratory environment, using light sources and detectors to directly measure the stray-light levels and distributions. It's capable of directly obtaining actual measured data with high reliability revealing practical issues overlooked by theoretical or simulation models. Nevertheless, under real complex conditions, the ray-tracing method cannot handle the computational burden of simultaneously considering numerous sources in the actual sky, while the experimental method is constrained by equipment and facility limitations, making it impractical to fully replicate the intricate environment of real space observations\citep{Zhang2023,Clermont2020}. Thus, we propose a strategy that separates the telescope’s inherent stray-light suppression performance from external illumination conditions. Specifically, we employ the Point Source Transmission (PST) to characterize CSST's stray-light suppression capability as a function of incident angle. PST is defined as the ratio of irradiance at the image plane from an off-axis point source to that from the same source positioned on-axis. Validated through simulations and laboratory experiments, out-of-field radiation entering the telescope is scattered by internal baffles and optical components, forming an approximately uniform stray-light background across the image plane \citep{WangWei2021}. Due to CSST’s asymmetrical structure, its PST is inherently a two-dimensional function. We obtained the corresponding PST matrix through a combination of experimental measurements and simulations. By integrating external illumination data (obtained from star catalogs) with the PST values at corresponding incident angles, we efficiently calculate off-field stray-light irradiance. 

In-field stray-light sources, such as zodiacal light and bright stars, are treated independently and then combined to form the total stray-light background. Among these, scattering and ghost images due to bright stars significantly impact imaging quality and exhibit relatively stable distribution patterns. Thus, evaluating their relative contributions is essential. 

In this work, we present a computational approach designed specifically for the CSST mission. By inputting the telescope’s orbital position, observational time, pointing direction, satellite attitude, relevant star catalogs, and parameters of mechanical and optical components, our model calculates the resulting stray-light background across the focal plane. Additionally, the distribution patterns of scattering and ghost images caused by bright in-field stars are derived. 

The remainder of this paper is structured as follows:  Section \ref{sect:The stray light sources and their propagations} describes the sources and propagation paths of stray light. Section \ref{sect:Stray light outside the field of view} presents the computational methods for off-field point sources and earthshine. Section \ref{sect:In FOV} elaborates on the computational methods for in-field zodiacal light, scattering, and ghost images. Finally,  Section \ref{sect:conclusion} summarizes our findings and discusses their implications for scientific image simulation and observational scheduling.


\section{STRAY LIGHT SOURCES AND THEIR PROPAGATION MECHANISMS }
\label{sect:The stray light sources and their propagations}
\subsection{Primary Sources}
In stray light analysis, most external sources—such as the Sun, Moon, major planets in the solar system, and bright stars—can effectively be approximated as point sources. Additionally, extended sources, including earthshine (Sun and Moon reflections from Earth’s surface) and zodiacal light, also contribute significantly to stray light. The typical spectral characteristics and intensities of these sources are summarized in Table~\ref{tab1}.
\begin{table}[htbp]
\bc
\begin{minipage}[]{100mm}
\caption[]{Characteristics of Stray Light Sources\label{tab1}}\end{minipage}
\begin{adjustbox}{width=\linewidth}
\small
 \begin{tabular}{ccccccc}
  \hline\noalign{\smallskip}
Source & Sun & Moon &Planet
& Bright Star& Space Background
& Earthshine\\
 &  &  &(Jupiter)& & (Zodiacal Light)& \\
  \hline\noalign{\smallskip}
Spectral &5700K Blackbody&Similar to Sun&Similar to Sun&Various &Similar to Sun&Similar to Sun\\
Distribution &Point Source&Point Source&Point Source&Point Source&Diffuse Background&Non-uniform Surface Source\\
Temporal &Sunlit Area&Visible Time&Visible Time&Visible Time&All Time&Earth bright edge\\
Magnitude&-26mag&-12mag(Full Moon)&-2.9mag&-1.5mag &~3×10$^{-18}$
erg/s/cm$^2$/A/arcsec$^2$&On Working Condition\\
  \noalign{\smallskip}\hline
\end{tabular}
\end{adjustbox}
\ec
\end{table}

According to the observational strategy, the Sun, Moon, and Earthshine are not permitted to appear within the field of view (FOV), so they exert influence from out of field (OOF). Planets within the solar system and bright stars generate stray light both from OOF and FOV, while their paths and mechanisms differ. Zodiacal light, as a full-sky background source, can enter the telescope simultaneously from both OOF and FOV. However, the portion of zodiacal light entering from OOF is greatly reduced by the telescope's inherent stray-light suppression structure, making its intensity significantly lower compared to the portion entering from within the FOV, typically 4-6 orders at small incident angles. From another aspect of view, the zodiacal light is also much fainter than the out-of-field stars, even we consider the total flux. Therefore, only the zodiacal light entering from within the FOV needs to be considered.

As the CSST’s main survey avoids the Galactic plane  and the bright stars are involved in the star catalog, galactic light contributions are excluded from this study. Additionally, airglow intensity across most CSST bands closely matches zodiacal light levels \citep{Leinert1998}. Given that CSST’s pointing directions maintain at least a 30° separation from Earth’s limb—where the PST value is approximately $10^{-6}$—stray light from airglow is negligible and thus omitted. 

\subsection{Out-of-field Stray Light Propagation Mechanisms}

\begin{figure}[H]
    \centering
    \includegraphics[width=0.7\linewidth]{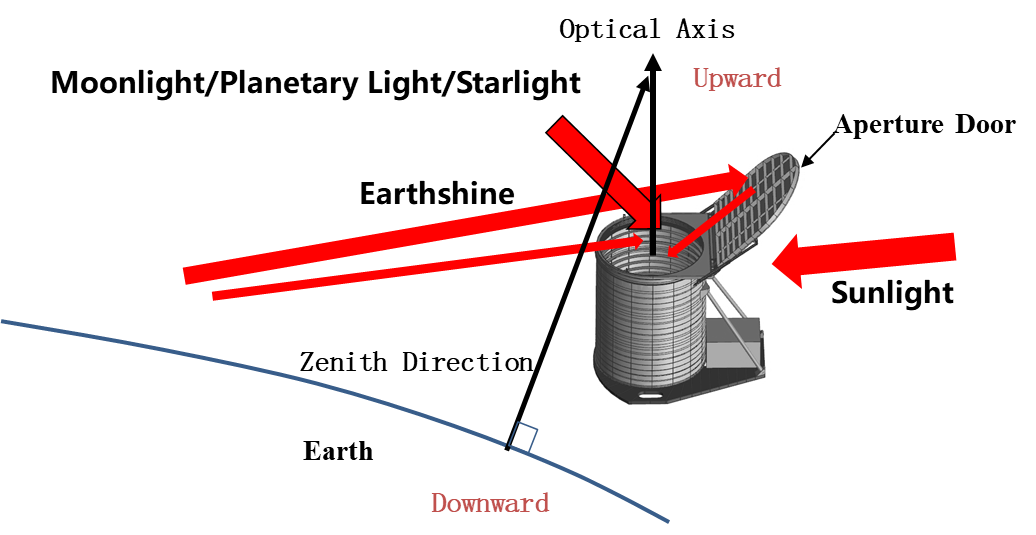}
    \caption{The main sources and propagations of stray light}
    \label{fig:The main sources and propagations of stray light}
\end{figure}

Internal structures of CSST, including lens hoods, blocking rings, and an aperture door coated with space-grade black paint, strongly suppress off-field stray light. Despite these measures, certain angles or conditions still permit stray light entry. Primary stray light sources and their propagation paths are detailed below (see also Fig.\ref{fig:The main sources and propagations of stray light}): 

1) Sunlight:

The Sun, as the brightest source, is typically blocked by the aperture door, which opens to 135° aligned with the solar azimuth, effectively blocking sunlight at angles beyond 65° from the optical axis. Under special circumstances, the door opens only to 105°, restricting sunlight entry at angles between 50° and 65°. Observations with the Sun within 50° of the optical axis are strictly prohibited.

2) Moonlight:

The intensity of the moonlight significantly varies with lunar phases. Ignoring slight deviations during lunar eclipses, lunar phases are quantified by the angular separation between the Moon-Earth and Sun-Earth lines within the ecliptic plane, being 180° at full moon. For intermediate phases, brightness ratios relative to the full moon condition are estimated proportionally. 

3) Planets:

Mercury’s and Venus’s phases are calculated similarly to lunar phases. Magnitudes for superior planets—Mars, Jupiter, Saturn, Uranus, Neptune, Pluto—are computed based on Earth-object distances: 
\begin{equation}
     E_i = E_0\cdot Ph\cdot (\frac{d_0}{d})^2
\end{equation}
Where $E_i$ is the irradiance of the celestial body at the telescope, $E_0$ is the irradiance of the celestial body at a reference distance $d_0$, $Ph$ is the phase coefficient of the celestial body, and $d$ is the actual object-telescope distance.

4) Bright Stars:

Positions of bright extra-solar-system stars change minimally over observational timeframes and thus are considered fixed. The bright stars considered in our model are taken from Gaia data release 2. Their corresponding Gaia magnitudes G, Bp, and Rp are first converted to Sloan Digital Sky Survey (SDSS) magnitude systems (refer to section 5.3.7 of the Gaia Data Release 2 Documentation) \citep{Katz2019}. We also use the Galaxia code \citep{Sharma2011} to generate a mock star sample with stellar properties (such as effective temperature, surface gravity, and metallicity) along with SDSS magnitudes. Stellar properties from the Galaxia sample are then assigned to the Gaia stars by matching their SDSS magnitudes. Using these stellar properties, a spectral energy distribution (SED) for each Gaia star is retrieved from a spectral template library based on the BT-Settl model \citep{Allard2011,Allard2014}. Given the SEDs, the SDSS magnitudes, and response curves of CSST and SDSS filters, we can then calculate the CSST magnitudes of these bright Gaia stars via synthetic photometry. 

Off-field stellar radiation enters the telescope aperture, subsequently scattering from baffles, diaphragms, and mirrors, finally contributing to stray light irradiance at the image plane.

5) Earthshine:

Earthshine originates from Earth’s bright regions illuminated by the Sun and dark regions illuminated by the Moon. Bright-region radiance depends on solar altitude angle and terrestrial reflectivity, whereas dark-region brightness mainly correlates with lunar phases. Thus, earthshine intensity significantly varies with orbital position.

Earthshine enters via two pathways:

\begin{itemize}
    \item When the angle between the telescope’s optical axis and zenith is less than 30° (optical axis oriented upward), earthshine predominantly enters from the bottom, reflecting first from the aperture door’s interior surface, subsequently scattering into the aperture.
    \item Conversely, at optical axis-zenith angles greater than 20°, terrestrial reflections may directly enter the aperture.
\end{itemize}
 
After entering the telescope, these out-of-field rays are scattered by internal baffles and optical components, forming a uniform stray-light background on the image plane.

\subsection{In-field Stray Light Propagation Mechanisms}

Aside from external sources, zodiacal light and bright stars within the field of view also contribute to stray light. Zodiacal light induces a nearly uniform stray-light background across the focal plane, dependent on ecliptic coordinates relative to the solar position. Bright in-field stars produce scattered and ghost images via optical surface irregularities and internal reflections.

\begin{figure}[H]
    \centering
    \includegraphics[width=0.9\linewidth]{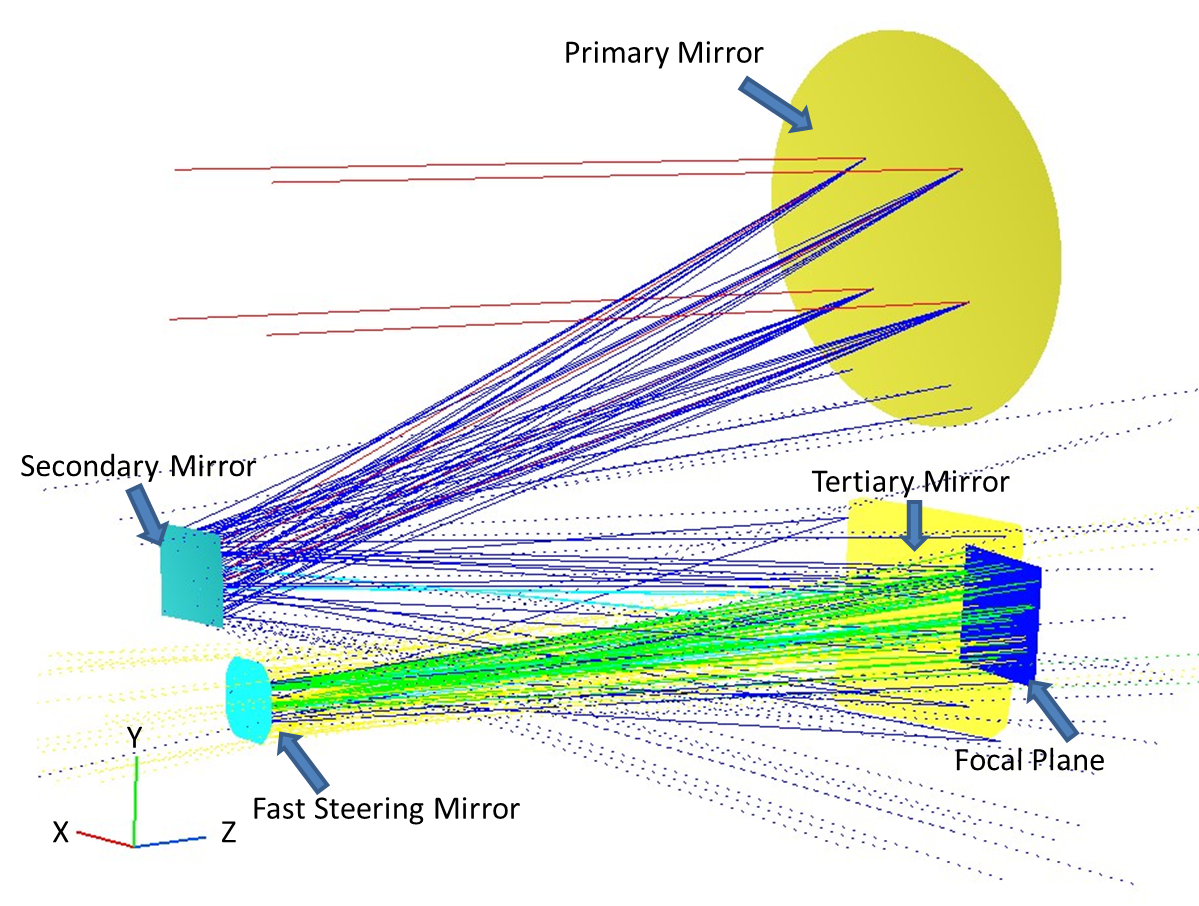}
    \caption{Scattering caused by CSST’s main system optical components (Primary mirror, Secondary mirror, tertiary mirror and fast steering mirror)
    We assume a single-direction parallel rays entering the system, which under ideal conditions would form a point image on the focal plane. However, due to scattering from the mirrors, stray light is distributed everywhere of the focal plane. }
    \label{fig:Scattering caused by CSST’s main system optical components}
\end{figure}

The main optical components of CSST include the primary mirror, secondary mirror, tertiary mirror, and fast steering mirror. Surface imperfections (irregularities, roughness) on these optical elements cause scattering. Ideally, parallel rays entering the telescope would form a perfect focal point, but surface scattering disperses stray light broadly across the focal plane, concentrating more strongly near image points (Fig. \ref{fig:Scattering caused by CSST’s main system optical components}.).

\begin{figure}[H]
    \centering
    \includegraphics[width=0.65\linewidth]{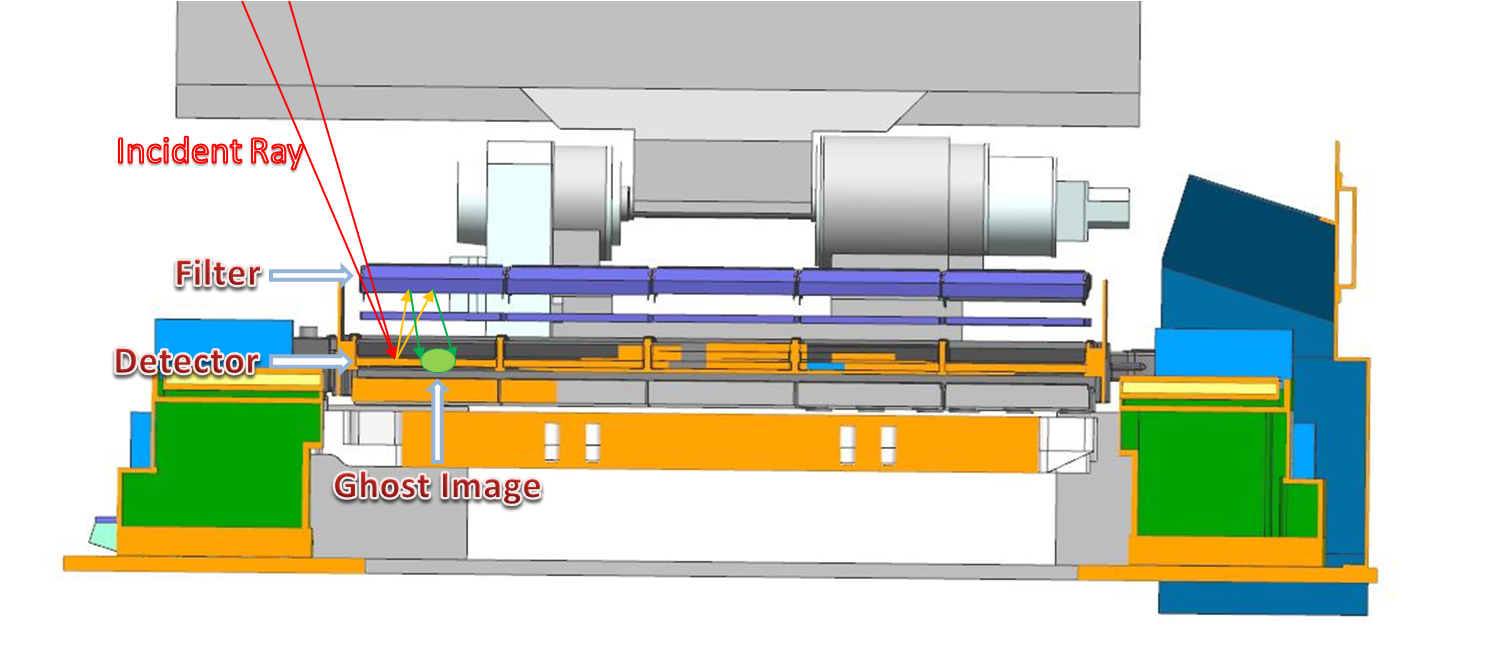}
    \caption{Reflections between filter and detector}
    \label{fig:Filter and Detector}
\end{figure}

CSST employs seven spectral bands for multi-band imaging, implemented using bandpass filters placed in front of detectors. Despite anti-reflective coatings, filters cannot achieve 100$\%$ transmission. Additionally, the detector surface, functioning effectively as a mirror, exhibits 20\% reflectivity. Multiple reflections between filters and detector surfaces consequently generate ghost images (Fig. \ref{fig:Filter and Detector}).

\section{Calculation methods for stray light from out-of-field sources}
\label{sect:Stray light outside the field of view}

\begin{figure}[H]
    \centering
    \includegraphics[width=1\linewidth]{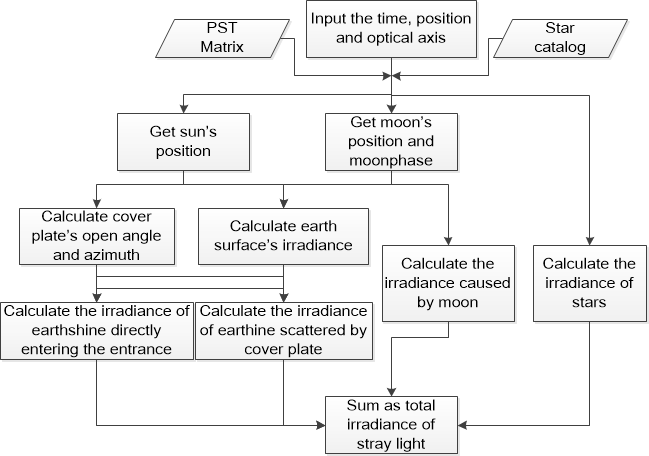}
    \caption{The calculation process of stray light from out-of-field sources}
    \label{fig:The calculation process of stray light outside the field of view}
\end{figure}

Off-field radiation entering the telescope is scattered by internal baffles and optical components, forming an approximately uniform stray-light background across the image plane \citep{WangWei2021}. This behavior has been validated through simulations and laboratory experiments. We adopt Point Source Transmission (PST) to quantitatively characterize the stray light transmission efficiency from the telescope entrance aperture to the image plane. By multiplying the irradiance of an out-of-field source at the entrance aperture by its corresponding PST value, the resultant stray light contribution can be accurately estimated. Fig.\ref{fig:The calculation process of stray light outside the field of view} illustrates the detailed calculation workflow. 

Due to CSST’s asymmetrical structure, PST depends not only on the angle $\theta$ between the incoming light source and the optical axis, but also on the azimuth angle $\phi$ relative to CSST’s orientation. Thus, the PST is inherently a two-dimensional function defined as: 
\begin{equation}
     PST(\theta,\phi) = \frac{E_d(\theta,\phi)}{E_i(\theta,\phi)}
\end{equation}
where $E_d$ is the stray light irradiance on the image plane and $E_i$ is the incident irradiance at the telescope’s aperture. 

To accurately derive the PST matrix for CSST, ray-tracing simulations alone are convenient but lack sufficient accuracy. Conversely, purely experimental measurements provide high accuracy but entail significant workloads. Therefore, we employ a hybrid methodology combining both approaches. First, we simulate PST values at various incident angles and azimuths using a detailed structural model of CSST. Subsequently, laboratory measurements are conducted on the actual CSST structure at representative angles using a collimated light source from a 2-meter aperture tube. By calibrating simulated results against these measured data points, we construct a reliable PST look-up matrix. 

\subsection{Calculation method for out-of-field point sources}

Stray-light irradiance from off-field point sources on the image plane is calculated by multiplying each source’s incident irradiance by its corresponding PST value and considering obstruction factors (whether the source is blocked by Earth or the aperture door).

Positions of the Sun, Moon, and solar-system planets vary over time. In our calculations, these positions are obtained from the Jet Propulsion Laboratory’s (JPL) DE405 ephemeris \citep{Standish1998}. By inputting the observation’s Julian date, we interpolate accurate celestial positions from this ephemeris.

For bright stars outside the solar system, relative positions remain nearly constant over typical observational periods. Star positions and magnitudes are thus directly obtained from the Gaia star catalog. Matching these Gaia stars with CSST’s simulation star catalog provides magnitudes in all seven CSST spectral bands: NUV(255$\sim$320nm), u(320$\sim$400nm), g(400$\sim$550nm), r(550$\sim$690nm), i(690$\sim$820nm), z(820$\sim$1000nm), y(940$\sim$1000nm).

Next, each source’s visibility (potential obstruction by Earth) is assessed. Two conditions must be simultaneously satisfied for a source to be obstructed by Earth:

1)  The angle between the source vector and telescope orbital-position vector (line connecting Earth’s center and telescope location) exceeds 90°.

2) The shortest distance between Earth’s center and the source-directed vector originating at the telescope position is less than Earth’s radius.

The aperture door position is solely determined by the Sun’s location, with a similar procedure used to calculate solar blockage.

Assume $n$ point sources, each with irradiance $E_i$, incident angle $\theta_i$, azimuth $\phi_i$, Earth-blockage factor $Se_i$, aperture-door blockage factor $Sp_i$, and corresponding PST value $PST(\theta_i,\phi_i)$. Total stray-light irradiance $E_p$ from these sources is:
\begin{equation}
     E_p=\sum_{i=1}^{n}E_iSe_iSp_iPST(\theta,\phi)
\end{equation}

\subsection{Calculation Method for Earthshine }

To model earthshine, Earth’s surface is subdivided into multiple discrete blocks, each approximated as a point source contributing individually to stray light. For each block, the angle and azimuth relative to CSST’s optical axis are denoted by $\theta_i$  and $\phi_i$, respectively, with corresponding irradiance $E_i$ reaching the telescope aperture. Considering aperture-door blockage $Sp_i$ and Earth blockage $Se_i$, the total irradiance from direct earthshine entering the aperture is calculated as: 
\begin{equation}
     EA_1=\sum_{i=1}^{n}E_iSe_iSp_iPST(\theta,\phi)
\end{equation}
Additionally, due to the aperture door’s proximity to the telescope aperture, some earthshine reflected off the aperture door may directly scatter into the telescope. Hence, it is necessary to consider the door-to-aperture scattering pathway. The aperture door, coated with black paint, is modeled as a Lambertian reflector with reflectance $r$. Dividing the aperture door and telescope aperture into multiple small discrete blocks, we calculate each block pair’s scattering contribution based on their relative geometrical relationships (incident angles $\theta_{ij}$, azimuth angles $\phi_{ij}$, and distances $d_{ij}$. The total irradiance from aperture-door scattered earthshine is: 
\begin{equation}
     EA_2=\sum_{i=1}^{m}MP_i\cdot r\cdot \sum_{j=1}^{k}\frac{PST(\theta_{ij},\phi_{ij})}{d_{ij}^2}
\end{equation}
where $Mp_i$ is the incident irradiance received by each door-block.

Thus, the total earthshine irradiance entering the telescope aperture is: $EA=EA_1+EA_2$.

\subsection{Comparision of Earthshine between CSST and HST}

To benchmark CSST’s earthshine levels against those experienced by Hubble Space Telescope (HST), we define a common observational scenario: at 00:50 UTC on January 1, 2026, both telescopes share identical right ascension and declination coordinates, situated within the sunlit region. For each telescope, earthshine irradiance is computed across an observational grid covering right ascension 0° to 345° and declination -75° to +75°. 

\begin{figure}[H]
    \centering
    \includegraphics[width=0.75\linewidth]{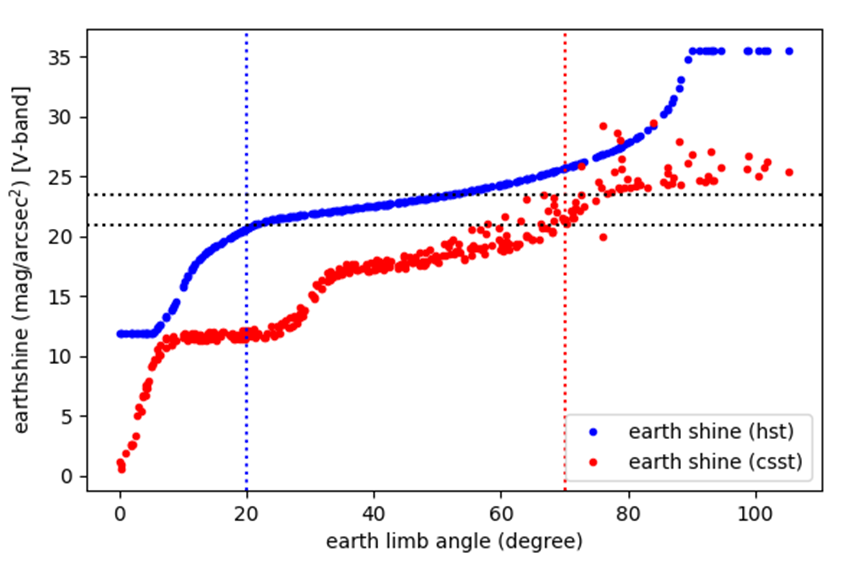}
    \caption{Comparison of Earthshine between CSST and HST in bright region}
    \label{fig:Comparison of Earthshine between CSST and HST}
\end{figure}

Fig.\ref{fig:Comparison of Earthshine between CSST and HST} illustrates these comparative results, indicating that CSST consistently experiences higher earthshine (combined effect of direct earthshine and earthshine scattered by the aperture door) irradiance than HST under identical observational conditions\footnote{The Earthshine brightness for HST is determined using the data presented in Figure 6.2 of the STIS Instrument Handbook (https://hst-docs.stsci.edu/stisihb/chapter-6-exposure-time-calculations/6-5-detector-and-sky-backgrounds)}. This discrepancy primarily arises from differences in internal telescope designs and orbital altitudes (CSST at approximately 400 km, versus HST at approximately 600 km). HST restricts observations within sunlit regions to angles greater than 20° from Earth’s bright limb. For CSST to achieve comparable earthshine conditions, observational angles must be maintained at least 70° from Earth's bright limb, a valuable insight informing CSST’s observational scheduling strategies. 

\section{Calculation methods for stray light from in-field sources}
\label{sect:In FOV}
In-field stray light primarily arises from zodiacal light and bright stars within the telescope's field of view. Zodiacal light generates a nearly uniform background across the focal plane, whereas bright stars induce scattering from optical surfaces and create ghost images due to reflections between the filter and detector. Each of these effects can be treated independently, and their combined results yield the total in-field stray light irradiance and distribution. 

\subsection{Calculation Method for Zodiacal Light }
Zodiacal light originates from sunlight scattered by interplanetary dust particles, predominantly concentrated near the ecliptic plane. To estimate its brightness, we reference observational brightness data at 0.5 µm wavelength presented by Leinert et al. (1998) \citep{Leinert1998}. Given an observational direction's ecliptic latitude and longitude difference relative to the Sun, zodiacal brightness at 0.5 µm is determined by interpolating tabulated observational data.

CSST’s main survey spectral coverage spans from 0.255 µm to 1 µm, similar to that of the Hubble Space Telescope (HST). By referencing HST's zodiacal spectral measurements, we calculate relative brightness ratios between 0.5 µm and other wavelengths. Using these ratios, the brightness at other wavelengths within CSST’s spectral range is estimated. Integrating over the wavelength interval yields the total brightness for the observational band:
\begin{equation}
     L=\int_{\lambda_0}^{\lambda_n}V(\lambda)d\lambda
\end{equation}
Here, $L$ is the total brightness over the observational spectral range, $\lambda_0$ and $\lambda_n$ are the lower and upper wavelength limits, and $V(\lambda)$ represents zodiacal brightness at wavelength $\lambda$.

Considering CSST’s entrance aperture diameter $D$ and focal length $f$, the resulting irradiance on the focal plane is:
\begin{equation}
     E=L\frac{\pi D^2}{4f^2} \label{eqution:Zodical}
\end{equation}
\begin{figure}[H]
    \centering
    \includegraphics[width=0.6\linewidth]{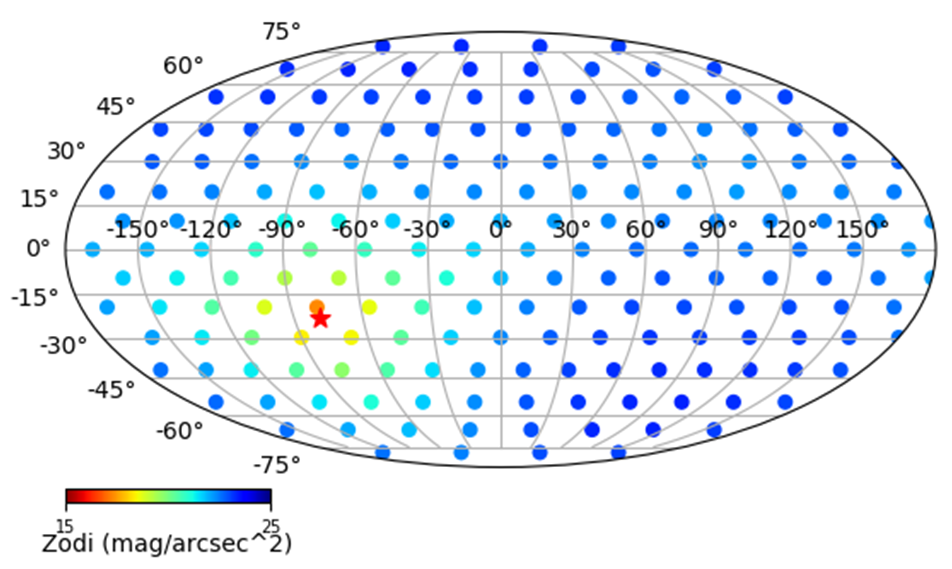}
    \caption{All-sky zodiacal light distribution in V-band (The red pentagram indicates the position of the Sun)}
    \label{fig:Zodiacal light distribution all-sky}
\end{figure}
Fig. \ref{fig:Zodiacal light distribution all-sky} shows the calculated zodiacal light distribution in the V-band across the full sky for January 1, 2026, at 00:50 UTC. The red star symbol indicates the solar position at that time. The computed distribution matches closely with observations from HST. 

\subsection{Scattering by Bright Stars on the Surfaces of Main Optical Components }
The Bidirectional Reflectance Distribution Function (BRDF) characterizes how incident radiation is scattered by an optical surface into various outgoing directions. BRDF essentially describes the distribution of reflected radiation as a function of incident and reflection angles, thus providing crucial data for stray light analysis.

Surface profiles of optical elements can be obtained using interferometry measurements, which are subsequently converted into Power Spectral Density (PSD) data. For optical surfaces with low roughness, PSD data can be reliably transformed into BRDF using Rayleigh-Rice perturbation theory. In practice, interpolating experimentally measured PSD data yields more accurate BRDF values for realistic scattering calculations.

In principle, the BRDF scattering model should incorporate the azimuthal dependence of both incident and scattered rays. However, this comprehensive treatment significantly increases computational complexity. Fortunately, precision-polished optical surfaces typically exhibit isotropic scattering behavior, meaning the BRDF mainly depends on the angle $\theta_s$ between the scattered ray and the specular reflection direction, and is nearly independent of azimuthal orientations.

For scattering from a single mirror surface, treated analogously to an equivalent refractive lens, the scattering radiance $L$, incident irradiance $E_i$, and BRDF are related as follows: 
\begin{equation}
     BRDF=\frac{L}{E_i}=\frac{\frac{d\Phi_s}{dA\cdot\cos\theta_sd\omega_s}}{\frac{d\Phi_i}{dA}}=\frac{d\Phi_s/(d(d_s^2)/f^2)}{d\Phi_icos\theta_s}=\frac{E_sf^2}{d\Phi_icos\theta_s}
\end{equation}
where $d\Phi_s$ and $d\Phi_i$ represent scattered and incident fluxes, respectively, $E_s$ denotes scattering irradiance, $f$ is the focal length, and $d_s$ is the distance between scattering point and image point. In optical systems with long focal lengths ($\theta_s\approx \frac{d_s}{f}$), the approximation $cos\theta_s\approx1$ holds true, simplifying the scattered irradiance calculation to: 
\begin{equation}
    E_s=\frac{d\Phi_iBRDF(\theta_s)}{f^2}
\end{equation}
Given the negligible contribution of higher-order scatterings, we limit our calculations to first-order scattering, thus considering each mirror's scattering independently and summing their contributions linearly.

Given BRDF of the four mirrors towards the detector's focal plane be denoted as $BRDF_t(\theta_s)$, and their clear aperture radii being $r_1, r_2, r_3, r_4$. The entrance pupil radius is denoted as $R$, the entrance pupil's scattering angle is $\theta_s$. The overall BRDF of the optical system can be expressed as:
\begin{equation}
    BRDF_t(\theta_s)=\frac{BRDF(\theta_s\frac{R}{r_1})}{r_1^2} + \frac{BRDF(\theta_s\frac{R}{r_2})}{r_2^2} + \frac{BRDF(\theta_s\frac{R}{r_3})}{r_3^2} + \frac{BRDF(\theta_s\frac{R}{r_4})}{r_4^2}
\end{equation}

Thus, the stray light irradiance distribution on the focal plane caused by scattering from a bright star is determined by: 
\begin{equation}
    E(x,y)=\frac{\pi R^2}{f^2}E_iBRDF_t(\theta_s) \label{equ:Scatter}
\end{equation}

The overall scattering stray-light distribution from multiple bright stars in the field can be computed individually using the above equation and subsequently summed together.

To verify the accuracy of our analytical method, a commercial optical simulation software (FRED) was employed for comparison. The simulations utilized the Harvey-Shack scattering model, described mathematically as \citep{Fest2013} :
\begin{equation}
    BRDF(\theta_s)=b_0[1+(\frac{sin\theta_s-sin\theta}{l})^2]^{\frac{s}{2}}
\end{equation}

\begin{figure}[H]
    \centering
    \includegraphics[width=0.9\linewidth]{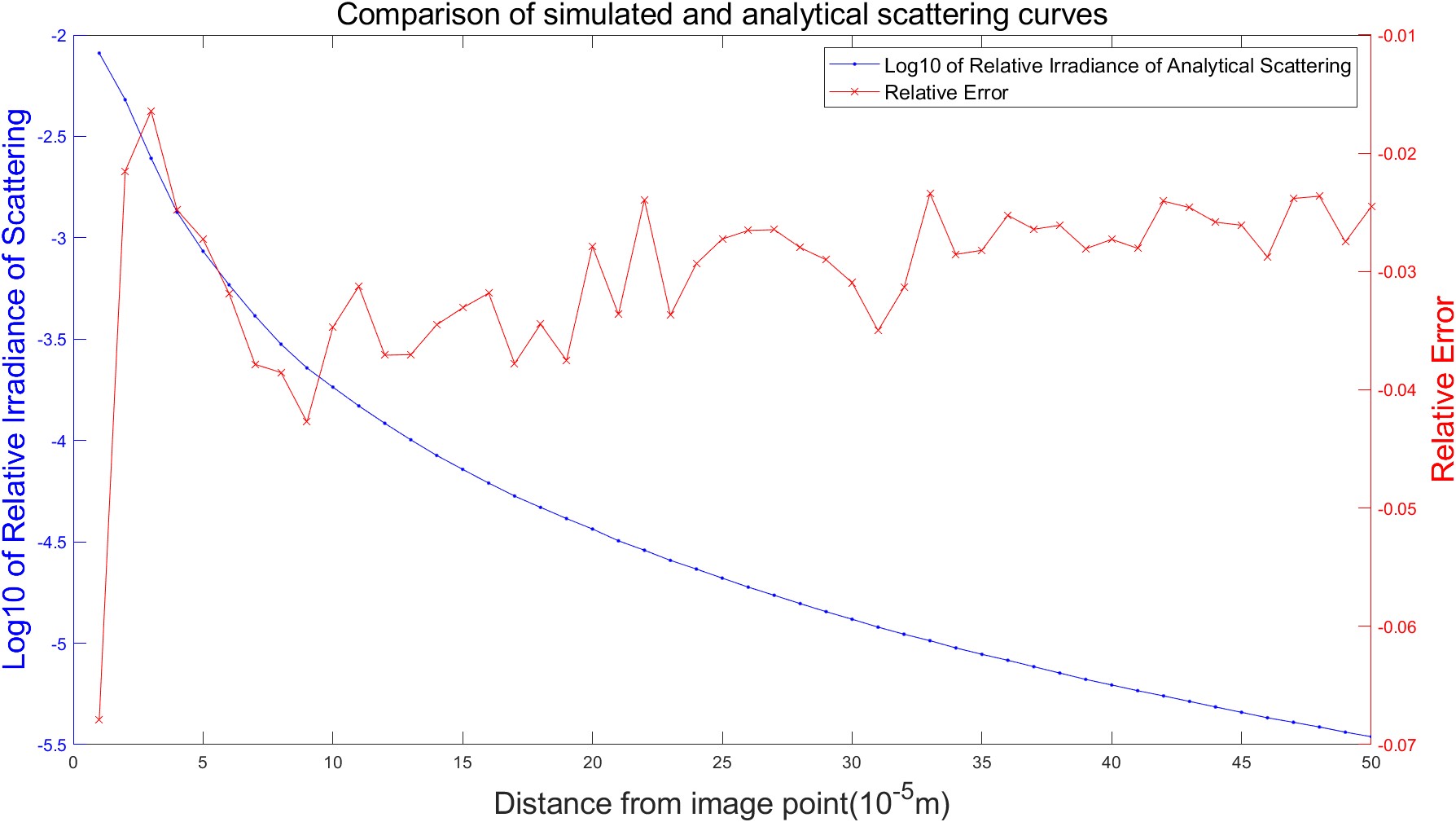}
    \caption{Comparison of simulated and analytical scattering curves}
    \label{fig:Comparison of simulated and analytical scattering curves}
\end{figure}

\begin{figure}[H]
    \centering
    \includegraphics[width=1\linewidth]{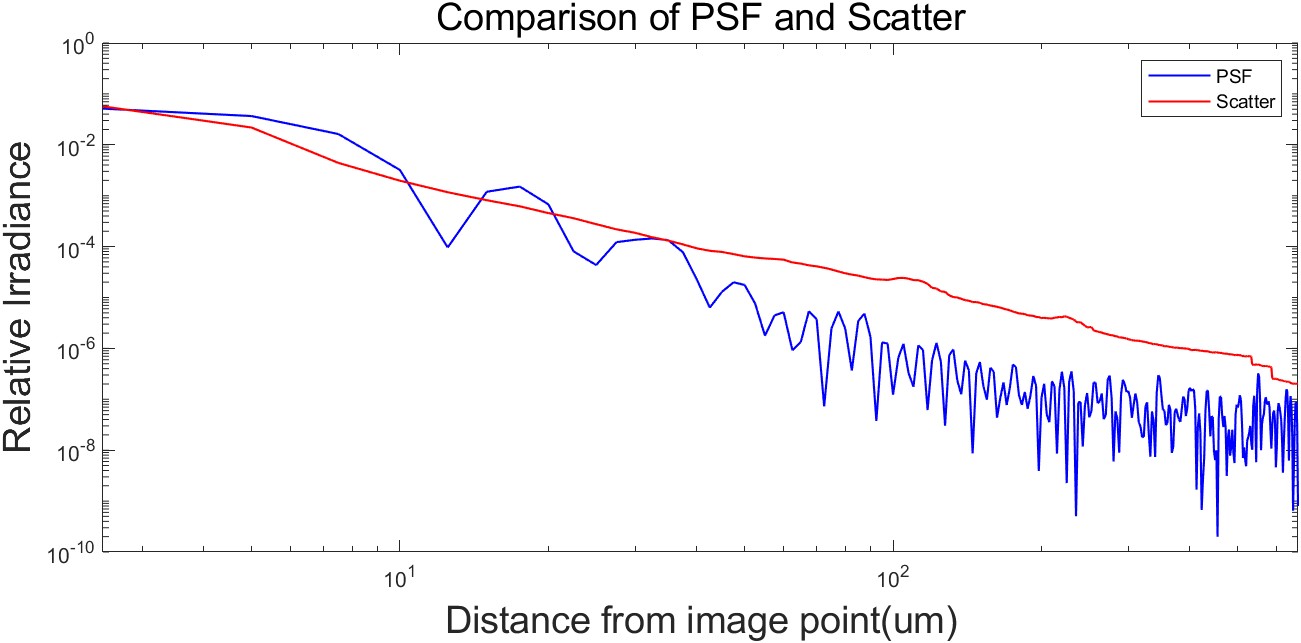}
    \caption{Comparison of PSF and Scattering curves}
    \label{fig:Comparison of PSF and Scatter}
\end{figure}

We adopted parameters $b_0=139257000, l=4e-6, s=-2.7$, representing a typical Total Integrated Scatter (TIS) of about $2\%$ for regular mirrors. Due to computational limitations, we compared the logarithm of the relative irradiance between the scattered point and image point within a radius of 50 pixels centered on the image position. Fig\ref{fig:Comparison of simulated and analytical scattering curves} shows the simulated and analytically computed scattering irradiance curves. The analytical results closely match the simulation, with deviations consistently less than 7\%, validating the accuracy and effectiveness of our scattering calculation algorithm. 
In practical CSST imaging calculations, measured mirror surface profiles obtained via interferometry are converted to pupil-plane amplitude and phase distributions. Through Fourier transformations, these data yield optical field distributions at the image plane. The sampling density of pupil-plane data directly influences the frequency coverage and accuracy of computed results. Typically, computational constraints necessitate relatively lower sampling density coupled with a two-dimensional Fourier transform to derive accurate point spread function (PSF) data in low-frequency spatial regions. However, increasing sampling density and utilizing one-dimensional Fourier transformations can precisely determine the isotropic scattering distribution in high-frequency regions. Fig.\ref{fig:Comparison of PSF and Scatter} illustrates a comparison between computed PSF curves and scattering distributions. 
Practically, results from PSF calculations are adopted in regions close to image points, while scattering distributions are employed for distant regions. Interpolation is used in transitional regions to ensure smooth continuity between PSF and scattering calculations. 

\subsection{Ghost Images Between the Filter and Detector}
Based on laboratory measurements, reflectance values of CSST filters across the seven imaging bands range from approximately $0.5\%$ to $3.8\%$. Each reflection thus significantly attenuates the energy transmitted through the system. Since ghost images can only form after an even number of reflections, we consider only second-order ghost reflections for our calculations, balancing accuracy and computational efficiency with actual measurement constraints. 

Each detector’s reflectance $R_d$ was estimated from measured Quantum Efficiency (QE) values using the relation $R_d\le {1-QE}$, as recommended by standard optical references and technical documents \citep{Fabricius2006,Howell2006}. We define $R_1$, $R_2$ and $R_d$ as reflectances of the filter’s front surface, filter’s back surface, and detector surface, respectively. Similarly, $T_1$ and $T_2$ represent the transmittance values of the filter’s front and back surfaces. 

\begin{figure}[H]
    \centering
    \includegraphics[width=0.8\linewidth]{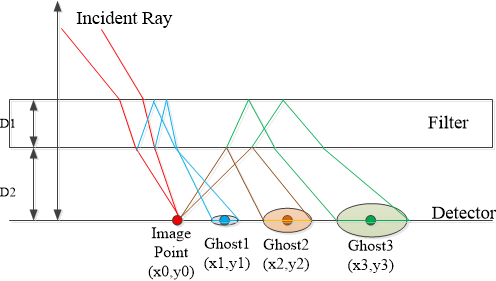}
    \caption{Ghost images caused by reflections between the filter and the detector}
    \label{fig:Ghost images}
\end{figure}

Considering these parameters, there are three possible second-order ghost image formation paths, illustrated in Fig.\ref{fig:Ghost images}, categorized according to their proximity to the nominal image position:

1) Reflection between the filter’s front and back surfaces;

2) Reflection between the filter’s back surface and the detector surface;

3) Reflection between the filter’s front surface and the detector surface.
The total energies associated with these ghost images ($ \Phi_1, \Phi_2, \Phi_3$) are calculated as follows:
\begin{equation}
      \Phi_1=\Phi T_1T_2R_1R_2,\quad \Phi_2=\Phi T_1T_2R_dR_2,\quad \Phi_3=\Phi T_1T_2^3R_1R_d
\end{equation}
Here, $ \Phi$ denotes the incident energy from the bright star.

The radii of these ghost images ($r_1, r_2, r_3$) are calculated using the geometry of the filter and detector, expressed as:
\begin{equation}
    r_1=2D_1\frac{u}{n},\qquad r_2=2D_2u,\qquad r_3=2u(\frac{D_1}{n}+D_2)
\end{equation}
where $D_1$ represents the distance from the filter's front surface to the filter's back surface, and $D_2$ represents the distance from the filter’s back surface to the detector, $u$ is the angle subtended by the entrance pupil rays on the detector, and $n$ is the refractive index of the filter material. 

Consequently, the irradiances ($E_1, E_2, E_3$) of these ghost images are given by: 
\begin{equation}
    E_1=\frac{\phi_1}{\pi r_1^2}=\frac{\Phi T_1T_2R_1R_2}{\pi (2D_1\frac{u}{n})^2},\quad 
    E_2=\frac{\phi_2}{\pi r_2^2}=\frac{\Phi T_1T_2R_dR_2}{\pi (2D_2u)^2},\quad 
    E_3=\frac{\phi_3}{\pi r_3^2}=\frac{\Phi T_1T_2^3R_1R_d}{\pi [2u(\frac{D_1}{n}+D_2)]^2}
\end{equation}
For CSST, the maximum angle between incident rays and the detector’s surface normal is typically no greater than 7°. Thus, small-angle approximations ($\theta=\sin(\theta)=tan(\theta)$) are justified. Assuming $\theta_x$ and $\theta_y$ represent ray inclination angles with respect to the detector’s $x$-axis and $y$-axis respectively, the central coordinates ($x_i,y_i$) of each ghost image relative to the nominal image point ($x_0,y_0$) can be expressed as: 
\begin{equation}
\begin{aligned}
    x_1=x_0+2D_1\frac{\theta_x}{n}&\quad y_1=y_0+2D_1\frac{\theta_y}{n}\\
    x_2=x_0+2D_2\theta_x&\quad y_2=y_0+2D_2\theta_y\\
    x_3=x_0+2D_1\frac{\theta_x}{n}+2D_2\theta_x&\quad y_3=y_0+2D_1\frac{\theta_y}{n}+2D_2\theta_y
\end{aligned}
\end{equation}
Finally, incorporating these positions, radii, and irradiances, the resulting ghost irradiance distribution function $E(x,y)$ on the detector can be summarized as: 
\begin{equation}
    E(x,y)=E_1circ(x_1,y_1,r_1)+E_2circ(x_2,y_2,r_2)+E_3circ(x_3,y_3,r_3)
\end{equation}
Here, $circ(x,y,r)$ denotes a circular function representing the geometric extent of each ghost image.

Taking the NUV band as an illustrative example, we computed the ghosting effects caused by an in-field bright star. We adopted an incident ray elevation angle of 3.9°, matching actual structural parameters, and set geometric parameters as $D_1=5mm, D_2=8mm, T_1=T_2=0.825, R_1=R_2=0.0093, R_d=0.15$.

\begin{figure}[H]
    \centering
    \includegraphics[width=0.9\linewidth]{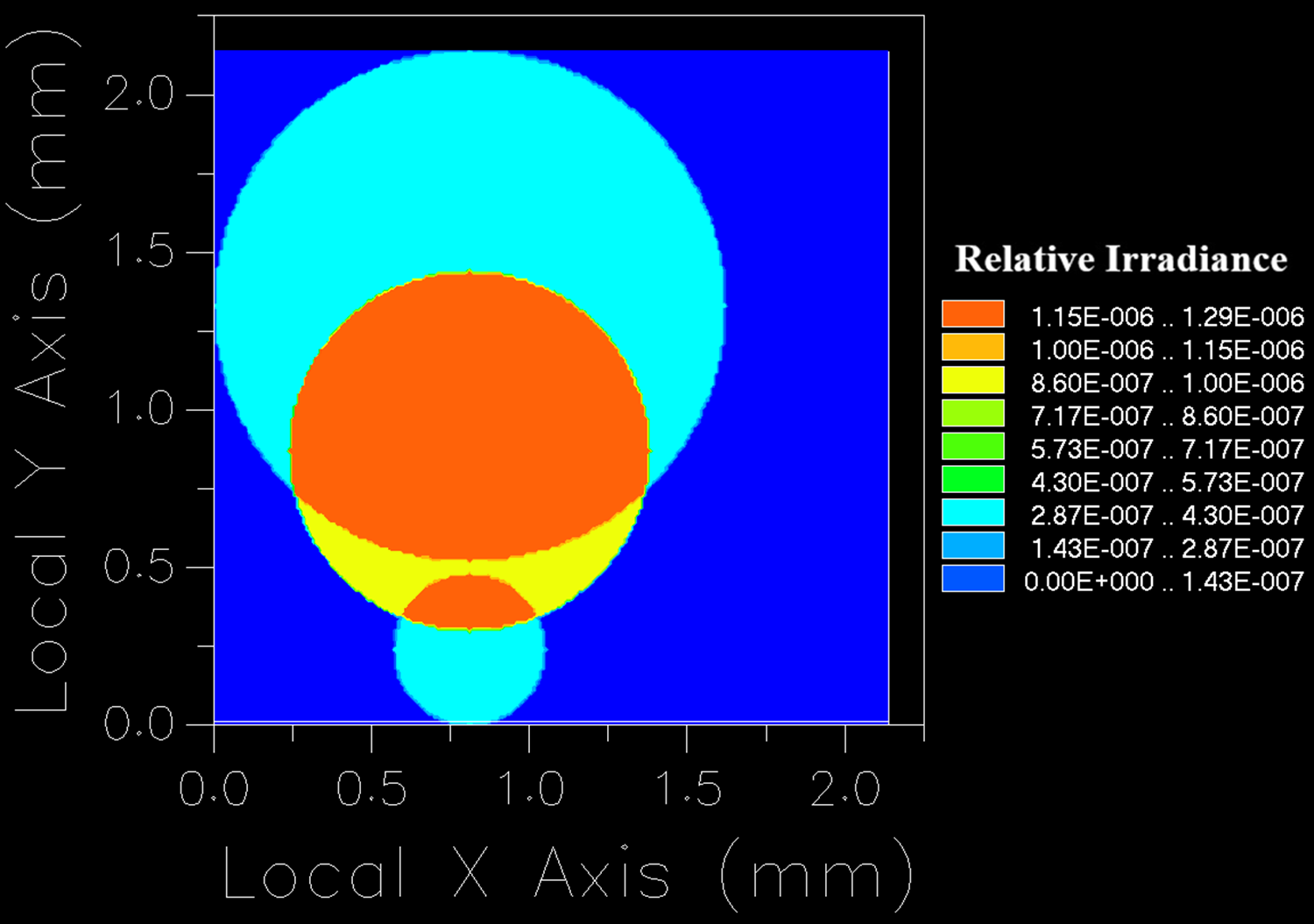}
    \caption{Ghost image ray tracing simulation and irradiance distribution}
    \label{fig:Ghost image ray tracing simulation and irradiance distribution}
\end{figure}

To validate these analytical calculations, results were cross-checked using the optical engineering simulation software FRED. Fig. \ref{fig:Ghost image ray tracing simulation and irradiance distribution} shows simulated ghost image positions and relative irradiance distributions. Simulation outcomes confirm that ghost images’ relative irradiances typically fall below $10^{-6}$ compared to the main stellar image, corroborating our analytical results. Similar agreement was observed across the other CSST imaging bands. 
In practical experiments using real devices, ghost images can be clearly observed by intentionally overexposing the sensor. By measuring the energies, positions, and sizes of these ghost images, and substituting these measured parameters into the derived equations, accurate determinations of the filter surfaces’ transmittance and reflectance, detector reflectance, and precise filter-detector distances can be achieved. Such experimental validations allow construction of highly accurate ghost image models for rigorous scientific image simulations. 

\subsection{Image simulation with in-field stray light}
\begin{figure}[H]
    \centering
    \includegraphics[width=0.8\linewidth]{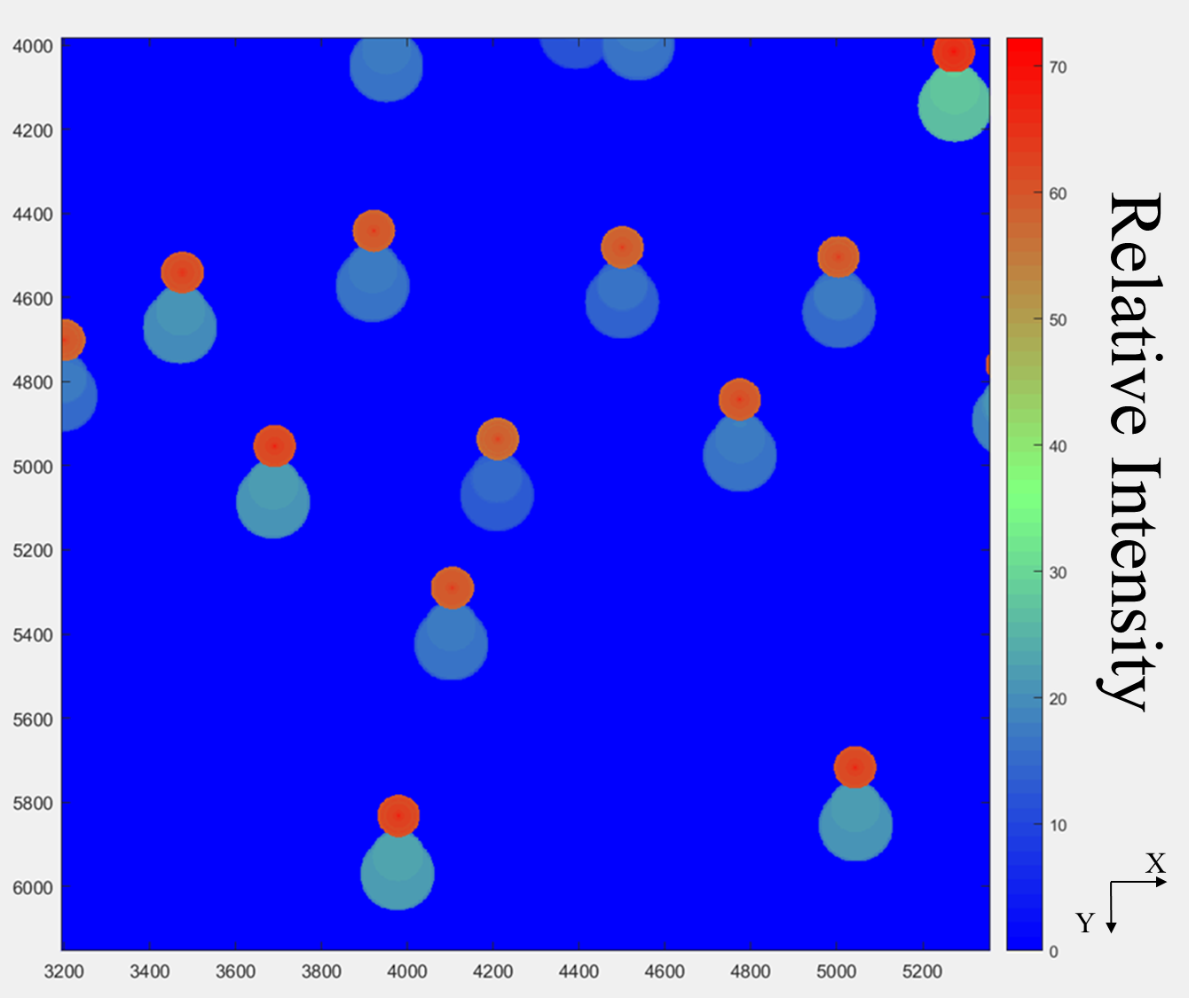}
    \caption{Scientific Simulation Image with In-Field Stray Light}
    \label{fig:Scientific Simulation Image with In-Field Stray Light}
\end{figure}

Integrating all in-field stray-light factors, we imported the simulated star catalog from the Cycle 9 data product of CSS-OS, generating scientific simulation multi-band images on the image plane. Fig\ref{fig:Scientific Simulation Image with In-Field Stray Light} shows an example from a partial area of one NUV-band detector.

In this figure, the reddish regions indicate the stellar image points and their scattering areas, while the cyan-colored regions represent ghost images. Data analysis reveals that the ghost images have a relative intensity of approximately $10^{-6}$ compared to the stellar image points. Regarding scattering effects, the relative intensity at a distance of 10 pixels from the stellar image points is about $10^{-4}$, and at a distance of 100 pixels, it drops to approximately $10^{-6}$.

\section{Conclusion}
\label{sect:conclusion}

This paper addresses the challenge of accurately assessing the level and spatial distribution of stray light in space-based astronomical telescopes under complex operational scenarios. We present a comprehensive calculation model that systematically incorporates all major sources of stray light, including sunlight, moonlight, major planets within the solar system, bright stars outside the solar system, earthshine, and zodiacal light.

Established on a robust logical framework, our model yields representative computational results for earthshine and zodiacal light under various realistic scenarios. By comparing these results with statistical data from the Hubble Space Telescope (HST), we demonstrate a consistent trend between the two, thus validating the accuracy of our method. Moreover, the comparative analysis highlights critical differences between the telescopes, providing valuable insights for astronomical observational scheduling.

Furthermore, the proposed model incorporates precise computational approaches to simulate both the scattering and ghost image distributions arising from bright stars in the field of view, ensuring that all dominant stray-light mechanisms are quantitatively accounted for. These simulation results serve as a critical reference for high-fidelity scientific image simulation and for understanding the impact of stray light on data quality.

The model has already been integrated into the CSST Main Survey Simulator. In future work, we will refine and calibrate the model parameters using actual measurement data obtained from engineering tests of the telescope, further improving its predictive accuracy and applicability in mission operations.

\normalem
\begin{acknowledgements}

This work was supported by National Astronomical Observatories, Chinese Academy of Sciences, under grant CSST-KSC-HTW-00-2023-009. L. L., X. Y.-H. and M. X.-M. acknowledge the science research grant from the China Manned Space Project with No. CMS-CSST-2021-B04.

\end{acknowledgements}

\bibliographystyle{raa}
\bibliography{bibtex}

\begin{thebibliography}{24}
\providecommand\natexlab[1]{#1}
\providecommand\JournalTitle[1]{#1}

\bibitem[Allard(2014)]{Allard2014}
Allard, F. 2014, Proceedings of the International Astronomical Union, 299, 271

\bibitem[Allard {et~al.}(2011)]{Allard2011}
Allard, F., Homeier, D., \& Freytag, B. 2011, 16th Cambridge Workshop on Cool Stars, Stellar Systems, and the Sun, 448, 91

\bibitem[Boyd {et~al.}(2022)]{Boyd2022}
Boyd, P.~T., Wilson, E.~L., Smale, A.~P., {et~al.} 2022, J. Astron. Telesc. Instrum. Syst., 8, 014003

\bibitem[Bruce(2015)]{Bruce2015}
Bruce, C.~C. 2015, Optical Engineering

\bibitem[Chabot {et~al.}(2023)]{Chabot2023}
Chabot, T., Brousseau, D., \& Thibault, S. 2023, Opt. Eng., 62, 025102

\bibitem[Clermont \& Michel(2024)]{Clermont2024}
Clermont, L., \& Michel, C. 2024, J. Appl. Rem. Sens., 18, 016508

\bibitem[Clermont {et~al.}(2020)]{Clermont2020}
Clermont, L., Michel, C., Blain, P., Loicq, J., \& Stockman, Y. 2020, Opt. Eng., 59, 025102

\bibitem[Fabricius {et~al.}(2006)]{Fabricius2006}
Fabricius, M., Bebek, C., Groom, D., Karcher, A., \& Roe, N. 2006, in Proceedings of SPIE, Vol. 6068

\bibitem[Fest(2013)]{Fest2013}
Fest, E. 2013, Stray Light Analysis and Control (Bellingham, Washington 98227-0010 USA: SPIE--The International Society for Optical Engineering)

\bibitem[Howell(2006)]{Howell2006}
Howell, S.~B. 2006, Handbook of {CCD} Astronomy, 2nd edn. (Cambridge University Press)

\bibitem[Johnson(2011)]{Johnson2011}
Johnson, B.~R. 2011, Stray Light Reduction in Optical Systems (SPIE)

\bibitem[Johnson(2017)]{Johnson2017}
---. 2017, Advanced Stray Light Analysis and Reduction Techniques (SPIE)

\bibitem[Kahan(2013)]{Kahan2013}
Kahan, M.~A. 2013, Journal of Astronomical Telescopes, Instruments, and Systems

\bibitem[Kahan(2019)]{Kahan2019}
Kahan, M.~A. 2019, Journal of Optical Technology

\bibitem[Katz {et~al.}(2019)]{Katz2019}
Katz, D., Sartoretti, P., Cropper, M., {et~al.} 2019, Astronomy \& Astrophysics, 622, A205

\bibitem[Krist {et~al.}(2023)]{Krist2023}
Krist, J.~E., Steeves, J.~B., Dube, B.~D., {et~al.} 2023, J. Astron. Telesc. Instrum. Syst., 9, 045002

\bibitem[Leinert {et~al.}(1998)]{Leinert1998}
Leinert, C., Bowyer, S., Haikala, L.~K., {et~al.} 1998, Astron. Astrophys. Suppl. Ser., 127, 1

\bibitem[Peterson(2004)]{Peterson2004}
Peterson, G.~L. 2004, in Proc. SPIE, Vol. 5178, Optical Modeling and Performance Predictions

\bibitem[Sharma {et~al.}(2011)]{Sharma2011}
Sharma, S., Bland-Hawthorn, J., Johnston, K.~V., \& et~al. 2011, The Astrophysical Journal, 730, 3

\bibitem[Smith(2000)]{Smith2000}
Smith, B.~A. 2000, Stray Light in Optical Systems (SPIE Press)

\bibitem[Standish(1998)]{Standish1998}
Standish, E.~M. 1998, {JPL} Planetary and Lunar Ephemerides, {DE405/LE405}, Interoffice Memorandum 312.F-98-048, JPL

\bibitem[WANG {et~al.}(2021)]{WangWei2021}
WANG, W., LU, L., ZHANG, T.-y., {et~al.} 2021, Chinese Optics, 14, 390

\bibitem[Zhan(2021)]{Zhan2021}
Zhan, H. 2021, Chinese Science Bulletin, 66, 1290

\bibitem[Zhang {et~al.}(2023)]{Zhang2023}
Zhang, E., Ye, W., Xia, Y., Wang, L., \& Zhang, M. 2023, Opt. Eng., 62, 034103

\end{thebibliography}

\end{document}